\UseRawInputEncoding
\documentclass{raa}            % referee version: for submission

\usepackage{graphicx,times}
\usepackage{natbib}
\usepackage{amssymb,amsmath}
\bibpunct{(}{)}{;}{a}{}{,}

\usepackage[pagebackref=true]{hyperref}
\hypersetup{colorlinks = true, linkcolor = green, anchorcolor = red, citecolor = blue, filecolor = red, pagecolor = red, urlcolor = red}

\begin{document}

   \title{Recurrent coronal jets observed by SDO/AIA}

   \volnopage{Vol.0 (20xx) No.0, 000--000} 
   \setcounter{page}{1}          

   \author{Yan-Jie Zhang
      \inst{1,2}
   \and Qing-Min Zhang
      \inst{1,2}
   \and Jun Dai
      \inst{1,2}
   \and Zhe Xu
      \inst{1}
   \and Hai-Sheng Ji
      \inst{1,2}
   }

   \institute{Key Laboratory of Dark Matter and Space Astronomy, Purple Mountain Observatory, CAS, Nanjing 210023, China; {\it zhangyj@pmo.ac.cn} \\
   \and
   School of Astronomy and Space Science, University of Science and Technology of China, Hefei, 230026, China
   \vs\no}
   %{\small Received~~20xx month day; accepted~~20xx~~month day}}

\abstract{In this paper, we carry out multiwavelength observations of three recurring jets on 2014 November 7. 
The jets originated from the same region at the edge of AR 12205 and propagated along the same coronal loop.
The eruptions were generated by magnetic reconnection, which is evidenced by continuous magnetic cancellation at the jet base.
The projected initial velocity of the jet2 is $\sim$402 km s$^{-1}$. The accelerations in the ascending and descending phases of jet2 are not consistent, the former is considerably larger than 
the value of $g_{\odot}$ at the solar surface, while the latter is lower than $g_{\odot}$. There are two possible candidates of extra forces acting on jet2 during its propagation. 
One is the downward gas pressure from jet1 when it falls back and meets with jet2. The other is the viscous drag from the surrounding plasma during the fast propagation of jet2.%The other is the aerodynamic drag force during the fast propagation of jet2.
As a contrast, the accelerations of jet3 in the rising and falling phases are constant, implying that the propagation of jet3 is not significantly influenced by extra forces.
\keywords{Sun:corona --- sun:activity --- sun:magnetic fields}
}

   \authorrunning{Zhang et al.}            
   \titlerunning{Recurrent coronal jets observed by SDO/AIA} 

   \maketitle

\section{Introduction} \label{sect:intro}
\hspace{1em} Jet-like motions at different scales are widespread in the solar atmosphere, including spicules \citep{holl82,dep07},  coronal jets \citep{Shimojo1996,cir07,ste15}, 
chromospheric jets \citep{nis11,tian18,wang21}, network jets \citep{Hassler1999,Tian2014}, and H$\alpha$ Surges \citep{Roy1973,Jiang2007,lihd17}. 
These phenomena are important to the transport of mass and energy toward the upper atmosphere and solar wind \citep{Brueckner1983}. 
Coronal jets were first observed by the Soft X-ray Telescope (SXT) on board the Yohkoh spacecraft \citep{Shibata1992} and have attracted a remarkable attention \citep{Raouafi2016,shen21}. 
With the unceasingly rapid development of solar telescopes, the understandings of coronal jets are greatly improved.
It is generally accepted that coronal jets are powered by magnetic reconnection, which occurs in the quiet region, coronal holes, and boundary of the active regions (ARs).
One of the observational evidences of magnetic reconnection is magnetic flux cancellation between the emerging dipolar flux and the ambient open fields or 
in converging magnetic system \citep{chae99,Chen2008,chen15,pan16,ste17}.

Coronal jets show diverse morphology. They are originally classified into the inverse-Y type and two-sided-loop type \citep{shibata1994,yoko95,tian17,shen19}.
According to the height of magnetic reconnection, they are also classified into the Eiffel tower type, $\lambda$ type, and micro-CME type \citep{Nistic2009}.
In addition, \citet{Moore2010} put forward the concept of blowout jets compared with the standard jets. 
The blowout jets are usually produced by the eruption of minifilaments \citep{Hong2011,li15,ste16,zhang2016}.
Some of the jets are related to the fan-spine topology with the presence of a magnetic null point \citep{wang2012,zhang2012,zhang2021,joshi20,yang20}.

Coronal jets are not always uniform in shape and temperature during their evolution. \citet{Pats2008} first reported helical structure in a coronal jet, which is verified to be a common 
feature \citep{Nistic2009,sch13,Zhang2014a,joshi18,chen21}. Both hot ($10^{5}$-$10^{6}$ K) and cold components ($10^{4}$-$10^{5}$ K) are found to coexist in coronal jets, although they are not strictly 
co-spatial \citep{ALEXANDER1999,Jiang2007,Kim2007,Nistic2011}. Using two-dimensional MHD numerical simulations, \citet{Nishizuka2008} reproduced the hot and cold components in jets.
Intermittent, bright and compact plasmoids (blobs) in recurring extreme-ultraviolet (EUV) jets were first observed and investigated by \citet{Zhang2014b}, 
which are explained as a result of tearing-mode instability in the current sheet where magnetic reconnection takes place. The sizes of plasmoids are extended to subarcsecond \citep{Zhang2019}
thanks to the high resolution of slit-jaw imager (SJI) on board the Interface Region Imaging Spectrograph \citep[IRIS;][]{de14}. 
\citet{li19} studied two coronal jets occurring during two M-class flares observed in 304 {\AA}. Many vortex-like structures are identified in the upstream and downstream regimes, 
which are interpreted by Kelvin-Helmholtz instability and a combination of Kelvin-Helmholtz instability and Raleigh-Taylor instability.

It is noted that not all jets escape the corona successfully. Some of them decelerate and fall back after reaching the apex.
For instance, \citet{Liu2009} investigated a chromospheric jet observed by Hinode on 2007 February 9 and obtained the mean projected acceleration (-141 m s$^{-2}$).
\citet{Zhang2014a} studied a swirling EUV jet related to a C1.6 flare on 2011 October 15 and calculated the mean apparent acceleration (-97 m s$^{-2}$).
\citet{Sakaue2018} analyzed a solar jet associated with a C5.4 class flare observed simultaneously in H$\alpha$ and soft X-ray on 2014 November 11.
The apparent acceleration (-176 m s$^{-2}$) and inclination angle ($\sim$50$^{\circ}$) of the jet were derived. 
It is concluded that the actual acceleration of the jet was consistent with the gravitational acceleration ($g_{\odot}\approx-274$ m s$^{-2}$) at the solar surface. 
In most cases, the acceleration values are basically the same in the ascending and descending phases of jets. When external forces apart from the gravity are involved, things are different. 
Using both spectroscopic and stereoscopic observations, \citet{lu19} studied a series of recurrent jets that fell back to the solar surface at speeds of 40$-$170 km s$^{-1}$ on 2014 July 7.
\citet{Huang2020} reported the multiwavelength observations of a solar jet propagating along a closed coronal loop on 2015 May 4.
Rapid change of the jet\rq{}s acceleration was found when the jet moved to the apex. It is proposed that the chromosphere evaporation at the remote footpoint propagates along the coronal loop 
and provides downward pressure on the jet, which causes the drastic change of acceleration.

Recurring jets are frequently observed \citep[e.g.,][]{chae99,chifor2008,chen15}. In this paper, we analyze recurrent jets containing three jets (jet1, jet2, and jet3) originating from the same base 
at the edge of AR 12205 on 2014 November 7. The acceleration of jet2 in the rising and falling phases show different values due to the influence of jet1. And jet3 serves as the control group. 
We describe the data analysis in Section~\ref{sect:data}. The results are represented in Section~\ref{sect:Results}. Comparison with previous works are discussed in Section~\ref{sect:discussion}.
A brief summary is given in Section~\ref{sect:sum}.

\section{Data analysis} \label{sect:data}
\hspace{1em} The data used here were observed by the Atmospheric Imaging Assembly \citep[AIA;][]{Lemen2012} and the Helioseismic and Magnetic Imager \citep[HMI;][]{Scherrer2012} 
on board the Solar Dynamics Observatory \citep[SDO;][]{Pesnell2012}. AIA takes full-disk EUV images in 94, 131, 171, 193, 211, 304, and 335 {\AA} with a time cadence of 12 s.
The photospheric line-of-sight (LOS) magnetograms are obtained from HMI with a cadence of 45 s. The level\_1 data of AIA and HMI are calibrated with a spatial resolution of 1.2$\arcsec$ 
using the standard Solar Software (SSW) programs \textit{aia\_prep.pro} and \textit{hmi\_prep.pro}, respectively.

\begin{figure}
\centering
\includegraphics[width=0.9\textwidth, angle=0]{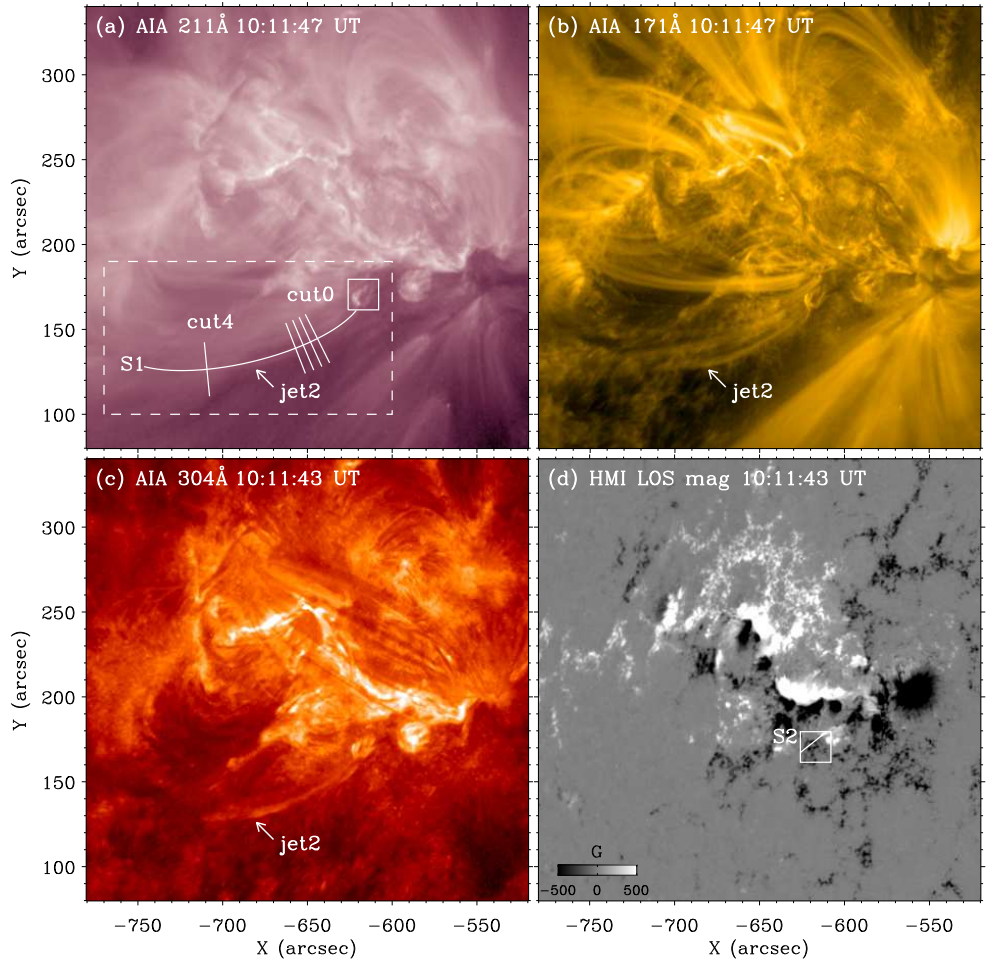}
\caption{(a-c) EUV images of AR 12205 observed by SDO/AIA in 211, 171, and 304 {\AA} during jet2. In panel (a), the dashed rectangle represents the field of view of Figure~\ref{Fig2}.
The white solid box delineates the footpoint of the recurring jets. A curved slice S1 is used to investigate the propagation of jets. The five short lines (cut0$-$cut4) are used to investigate the possible
untwisting motion of jets.
(d) Photospheric LOS magnetogram of AR 12205 observed by SDO/HMI. The solid box has the same meaning as that in panel (a).
A short line S2 is used to investigate the evolution of LOS magnetic field.}
\label{Fig1}
\end{figure}

\section{Results} \label{sect:Results}
\begin{figure}
\centering
\includegraphics[width=0.9\textwidth, angle=0]{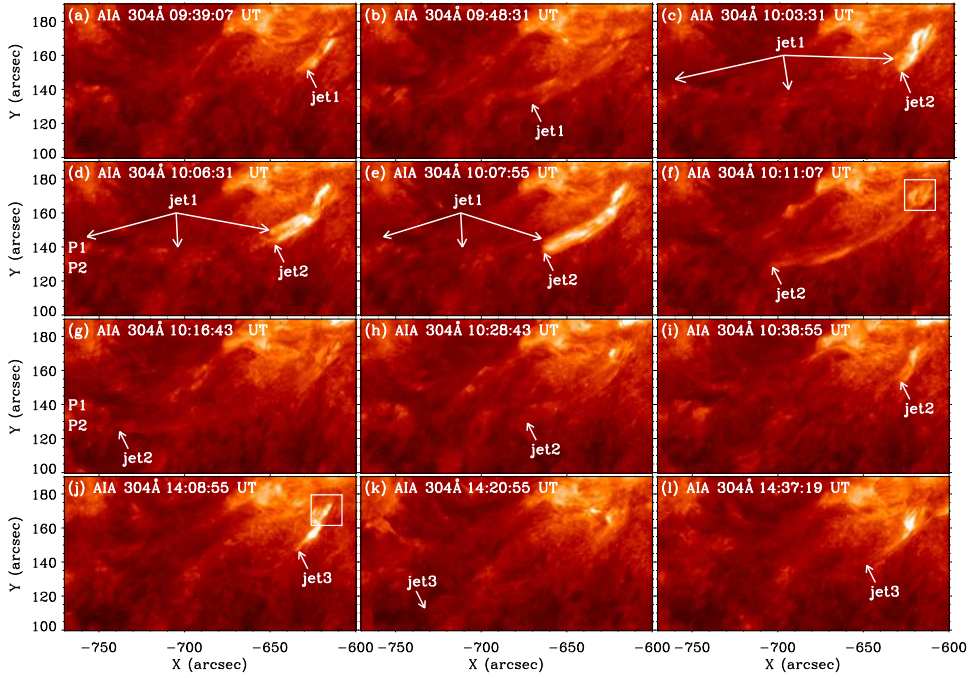}
\caption{Twelve snapshots of the AIA 304 {\AA} images during 09:39$-$14:37 UT. The white arrows point to jet1, jet2 and jet3. P1 and P2 in (d) and (g) illustrate the position of flux tube.
The white boxes in (f) and (j) represent the footpoint of jets.}
\label{Fig2}
\end{figure}

\hspace{1em} Figure~\ref{Fig1}(a-c) show EUV images of AR 12205 observed by AIA in 211, 171, and 304 {\AA} during jet2, respectively.
The dashed rectangle in panel (a) represents the field of view (FOV) of Figure~\ref{Fig2}. The base of the jets is within the white solid box. 

In Figure~\ref{Fig2}, twelve snapshots of the AIA 304 {\AA} images illustrate the temporal evolution of the recurring jets, which belong to the blowout type due to their visibility in 304 {\AA}.
The jet1 appeared at $\sim$09:38 UT and propagated along the guiding magnetic field as is shown in Figure~\ref{Fig2}(a-b). 
At $\sim$10:02 UT, the jet2 started to erupt, when jet1 was on its way falling down (see Figure~\ref{Fig2}(c)). 
The jet1 continued to move downward along the flux tube when jet2 keeps rising, until it merged into the jet2 at $\sim$10:10 UT (see Figure~\ref{Fig2}(d-f)). 
The jet2 fell back to the footpoint at $\sim$10:38 UT. The positions of flux tube at 10:06:31 UT and 10:16:43 UT are indicated by P1 and P2 in Figure~\ref{Fig2}(d) and (g), respectively.
The transverse drift of the flux tube may suggest that the jet2 did not follow the same path during its ascending and descending phases. 

During 10:38$-$14:08 UT, a few jets occurred from the jet footpoint, but with much smaller temporal and spatial scales, which are out of the scope of this study.
The jet3 appeared at $\sim$14:08 and reached its apex at $\sim$14:20 UT, and finally fell back to the footpoint at $\sim$14:37 UT (see Figure~\ref{Fig2}(j-l)).
It is seen from the online animations (\textit{Fig2-1.mp4}, \textit{Fig2-2.mp4}) that the jets in four passbands (171, 193, 211, and 304 {\AA}) were roughly cospatial and cotemporal.    

\begin{figure}
\centering
\includegraphics[width=0.9\textwidth, angle=0]{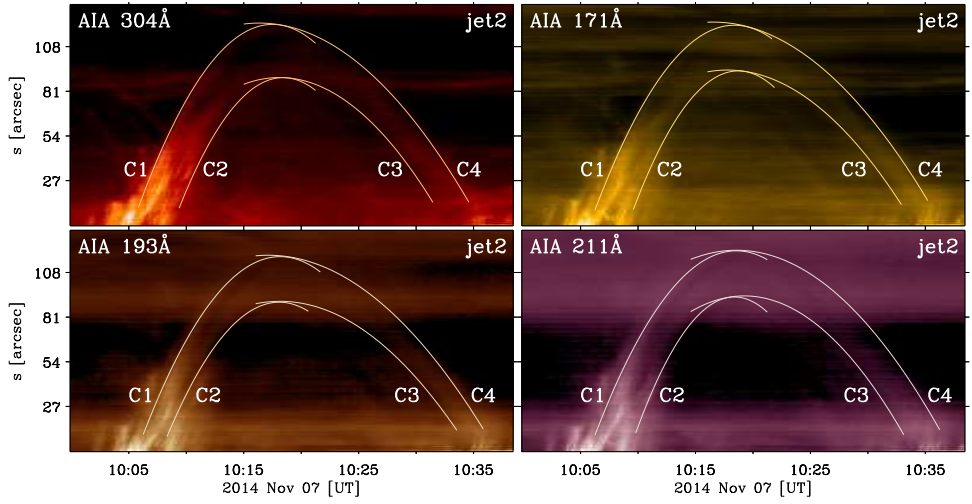}
\caption{Time-distance diagrams of slit S1 in 304, 171, 193, and 211 {\AA} during jet2.
The solid curves C1 and C2 represent the outer and inner boundaries of jet2 in the rising phase,
while the curves C4 and C3  represent the outer and inner boundaries of jet2 in the falling phase.}
\label{Fig3}
\end{figure}

In order to investigate the longitudinal motion of jet2, we draw a curved slit S1 along the axis of jet, which is 134$\arcsec$ in length and 10$\arcsec$ in width (see Figure~\ref{Fig1}(a)).
The time-distance diagrams of S1 in 304, 171, 193 and 211 {\AA} during jet2 are displayed in Figure~\ref{Fig3}. 
The ascending and descending motions of jet2 are clearly illustrated by the parabolic trajectory.
The solid curves C1 and C2 represent the outer and inner boundaries of the trajectory during the ascending phase, 
while the curves C4 and C3 represent the outer and inner boundaries of the trajectory during the descending phase.
They are fitted with a quadratic function \citep{Zhang2014a}:
\begin{equation} \label{eqn1}
    s(t) = \frac{1}{2}a_0(t-t_0)^2 + b_0(t-t_0) + c_0,
\end{equation}
where $t_0=10:00$ UT, $b_0$ denotes the projected initial velocity at $t_0$, $a_0$ denotes the projected acceleration of jet2 at the plane of sky.

The results of curve fitting for C1 and C2 are used to obtain the initial velocity and acceleration during the ascending phase.
The mean values of $b_0$ and $a_0$ in the four passbands are listed in second and third column of Table~\ref{Tab1}.
Likewise, the results of curve fitting for C4 and C3 are used to obtain the acceleration during the descending phase.
The mean values of $a_0$ in the four passbands are listed in the fourth column of Table~\ref{Tab1}.
It is seen that the projected initial velocity of jet2 lies in the range of 360$-$430 km s$^{-1}$ and has an average of $\sim$402 km s$^{-1}$.
The acceleration of jet2 during the rising phase, even taking into account of projection effect, is much greater than $g_\odot$,
while the acceleration during the falling phase is significantly less than $g_\odot$.

\begin{table}
\begin{center}
\caption[]{Mean Value of Initial Velocity ($b_0$) and Acceleration in the Ascending ($a_{rise}$) and Descending ($a_{fall}$) Phases of Jet2 in the four EUV Passbands.}
\label{Tab1}
 \begin{tabular}{cccc}
  \hline\noalign{\smallskip}
passband & $b_0$  &  $a_{rise}$  &  $a_{fall}$                     \\
    ({\AA})    & (km s$^{-1}$)    &  (m s$^{-2}$)    &  (m s$^{-2}$)           \\
  \hline\noalign{\smallskip}
304  & 401  & -388 & -164   \\
171  & 412  & -385 & -145   \\
193  & 366  & -351 & -138   \\
211  & 428  & -405 & -171   \\
\hline
ave. & 402 & -382 & -155  \\
  \noalign{\smallskip}\hline
\end{tabular}
\end{center}
\end{table}

\begin{figure}
\centering
\includegraphics[width=0.9\textwidth, angle=0]{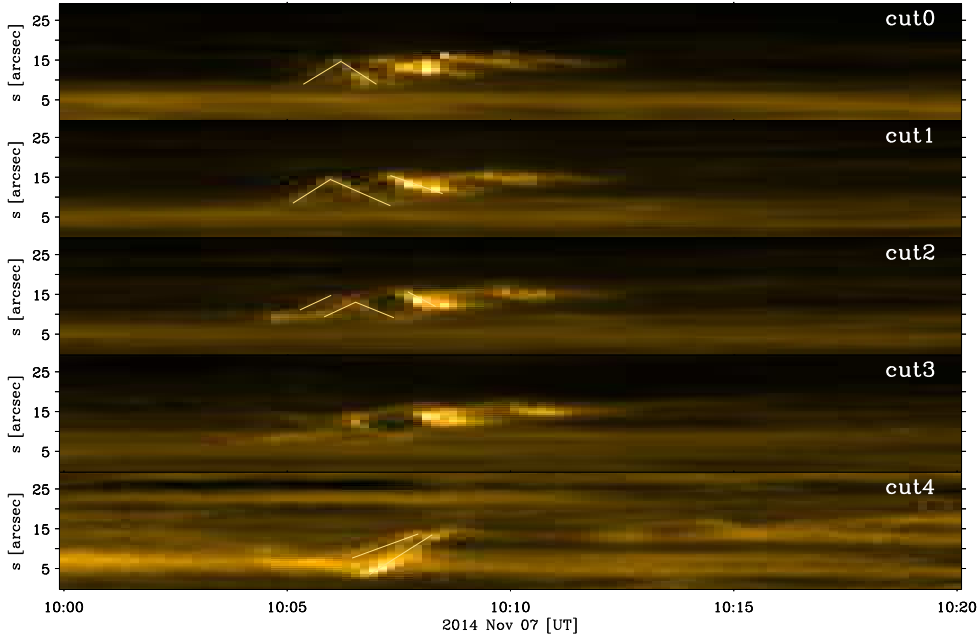}
\caption{Time-distance diagrams of the short slits cut0$-$cut4 in 171 {\AA}.
The solid lines are linear fittings of the transverse motion of jet2.}
\label{Fig4}
\end{figure}

In order to explore the transverse motion of jet2, we take five narrow slits (cut0$-$cut4) across the axis in Figure~\ref{Fig1}(a), which are 32$\arcsec$ in length and 5$\arcsec$ in width.
In Figure~\ref{Fig4}, the time-slice diagrams of cut0$-$cut4 in 171 {\AA} illustrate the transverse motion of jet2. 
The diagrams of cut0$-$cut2 show the transverse oscillation motion of the jet2 at the onset of eruption, and the diagram of cut3 reveals the disappearance of the transverse motion %oscillation
when the jet rose to a certain height, indicating that the rotation weakened as the jet moved upward.
The apparent velocities of the transverse oscillation are 10$-$20 km s$^{-1}$ using a linear fitting (see the solid lines).
The transverse rotation of the jet during its rising phase is similar to the event on 2011 October 15 \citep{Zhang2014a}, which is interpreted as the impulsive release and transfer of the 
accumulated magnetic helicity \citep{fang14}. 

\begin{figure}
\centering
\includegraphics[width=0.9\textwidth, angle=0]{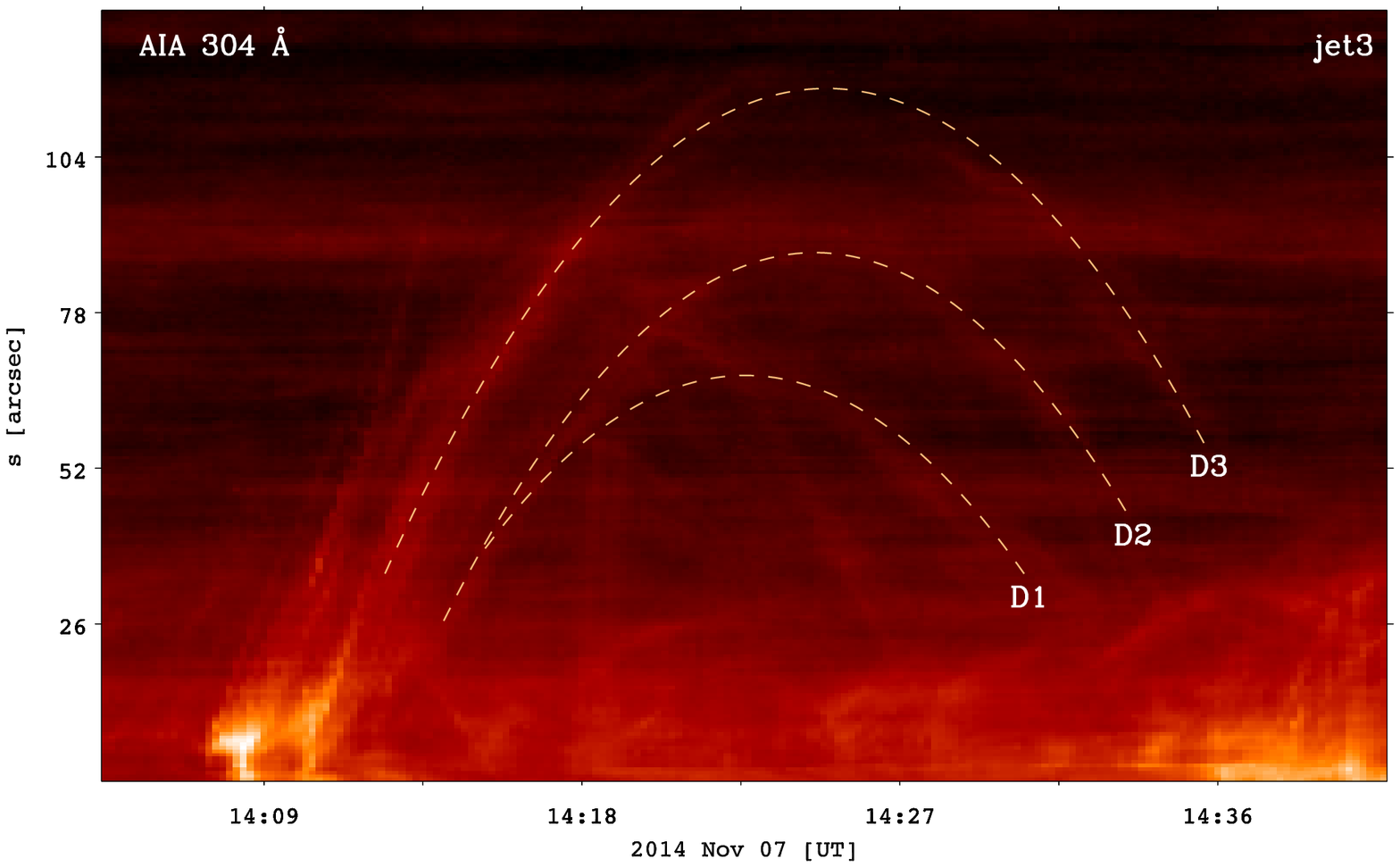}
\caption{Time-distance of S1 in 304 {\AA} during jet3. Three curves (D1, D2, and D3) represent the trajectories of three segments of jet3.}
\label{Fig5}
\end{figure}

Between jet2 and jet3, there were small-scale and short-lived jets originating from the jet base, which are out of the scope of this article.
The time-distance diagram of S1 in 304 {\AA} during jet3 is shown in Figure~\ref{Fig5}, showing its longitudinal motion. Note that jet3 could not be clearly observed in other passbands.
Three segments (D1, D2 and D3) could be distinctly identified in the diagram, which are outlined with dashed lines and fitted with a quadratic function:
\begin{equation} \label{eqn2}
   s(t) = \frac{1}{2}a_1(t - t_1)^2 + b_1(t - t_1) + c_1,
\end{equation}
where $t_1=13:50$ UT, $b_1$ denotes the projected initial velocity at $t_1$, and $a_1$ denotes the acceleration. The results of curve fitting for the three segments are listed in the Table~\ref{Tab2}.
It is shown in Figure~\ref{Fig5} that the trajectories of the segments of jet3 can be well described with parabolic curves of constant acceleration in the ascending and descending phases.
The initial velocities with an average value of $\sim$250 km s$^{-1}$ are lower than those of jet2. The average acceleration is $\sim$210 m s$^{-2}$, which is lower than $g_{\odot}$.

\begin{table}
\begin{center}
\caption[]{Fitted Values of Initial Velocity ($b_0$) and Accelerations in the Ascending ($a_{rise}$) and Descending ($a_{fall}$) Phases of the three segments of Jet3.}
\label{Tab2}
 \begin{tabular}{cccc}
  \hline\noalign{\smallskip}
segment & $b_1$  & $a_{rise}$    &    $a_{fall}$                   \\
               &  (km s$^{-1}$)    &  (m s$^{-2}$)      &   (m s$^{-2}$)            \\
  \hline\noalign{\smallskip}
D1  & 252  & -203 & -203   \\
D2  & 270  & -218 & -218   \\
D3  & 230  & -208 & -208   \\
\hline
ave. & 250 & -210 & -210 \\
  \noalign{\smallskip}\hline
\end{tabular}
\end{center}
\end{table}

\begin{figure}
\centering
\includegraphics[width=0.9\textwidth, angle=0]{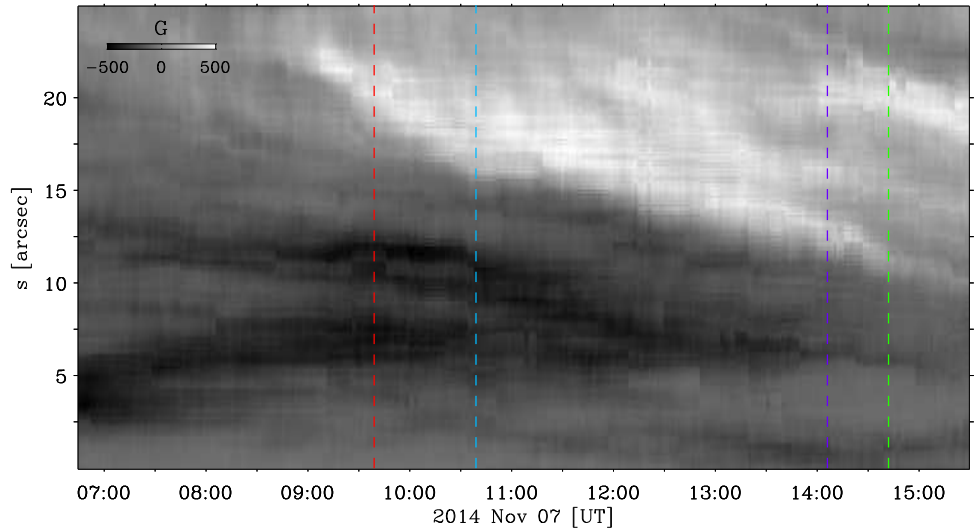}
\caption{Time-distance diagram of S2, showing the magnetic cancellation at the jet base.
The red, blue, purple, and green dashed lines signify the start time of jet1, end time of jet2, start time of jet3, and end time of jet3, respectively.}
\label{Fig6}
\end{figure}

To investigate the triggering mechanism of the recurrent jets, we draw the photospheric LOS magnetogram observed by HMI at 10:11:43 UT in Figure~\ref{Fig1}(d).
Like in Figure~\ref{Fig1}(a), the solid box signifies the area of jet base. A short slit (S2) within the box is used to investigate the magnetic field evolution.
The time-distance diagram of S2 is displayed in Figure~\ref{Fig6}. The red, blue, purple, and green dashed lines signify the start time of jet1, end time of jet2, start time of jet3, 
and end time of jet3, respectively. It is obvious that the positive and negative polarities were continuously approaching each other during the recurrent jets, 
implying that the three jets were generated by magnetic reconnection as a result of magnetic flux cancellation \citep{pan16,ste17,2017Chen}.

\section{Discussion} \label{sect:discussion}
\hspace{1em} As mentioned in Section~\ref{sect:intro}, the falling back of solar jets after rising to their apex is not uncommon. 
The acceleration usually keep constant during the whole evolution, implying a free fall.
\citet{Liu2009} reported recurrent chromospheric jets on the west limb observed in Ca {\sc ii} passband ($\log T\approx4.0-4.3$). The jets lasted for more than one hour with intervals of 12$-$14 minutes.
The average projected acceleration is $\sim$-141 m s$^{-2}$ (see Table~\ref{Tab3}).
\citet{Zhang2014a} analyzed a swirling flare-related coronal jet occurring at the edge of AR 11314. The projected acceleration of the jet is calculated to be -97 m s$^{-2}$.
\citet{Sakaue2018} studied a solar jet related to a C5.4 flare. The projected acceleration (-176 m s$^{-2}$) and inclination angle ($\sim$50$^{\circ}$) of the jet are derived.
It is found that the projected accelerations in the previous works are lower than $g_{\odot}$. The difference in values is mostly attributed to the different inclination angles. 

\begin{table}
\begin{center}
\caption[]{Comparison of the Mean Values of Accelerations in the Ascending ($a_{rise}$) and Descending ($a_{fall}$) Phases with Previous Works.}
\label{Tab3}
 \begin{tabular}{cccc}
  \hline\noalign{\smallskip}
event &   $a_{rise}$     &    $a_{fall}$      & Ref.           \\
           &  (m s$^{-2}$)                      &  (m s$^{-2}$)          &            \\
  \hline\noalign{\smallskip}
2007/02/09     &           -141                  &             -141           &       \citet{Liu2009}          \\
2011/10/15     &            -97                   &              -97            &      \citet{Zhang2014a}                  \\
2014/11/11     &           -176                  &             -176            &  \citet{Sakaue2018}               \\
2015/05/04    &   0$-$1000  &     1500$-$3000     &         \citet{Huang2020}      \\
       jet2            &           -382                  &              -155           &        this study                \\
       jet3            &           -210                  &              -210           &        this study               \\
  \noalign{\smallskip}\hline
\end{tabular}
\end{center}
\end{table}

However, when external forces are involved, the acceleration of a jet changes. \citet{Huang2020} reported the observation of a jet propagating along a closed coronal loop. 
Instant brightening is found at the remote footpoints of the loop, which is probably heated by the nonthermal electrons, MHD waves, and/or a conduction front generated by 
the magnetic reconnection associated with the jet. Extension of brightening along the loop, which is interpreted by chromospheric evaporation, meets with the ejecta near the loop apex and 
acts as a brake on the ejecta, leading to a strong deceleration (see Table~\ref{Tab3}).

In this work, we study three recurring jets originating from the same region and propagating along the same coronal loop. The first two jets erupted with short time interval, while the third jet occurred independently. As jet2 rises at an initial velocity of $\sim$400 km s$^{-1}$, it shows unwinding motion, which implies the release and transfer of magnetic helicity.
Interestingly, the accelerations in the ascending ($a_{rise}$) and descending ($a_{fall}$) phases of jet2 are not the same (see Table~\ref{Tab1}). 
The mean value of $a_{rise}$ is larger than $g_{\odot}$, suggesting that additional force is at work apart from the gravity.
Considering the short interval between jet1 and jet2 and the fact that jet1 moved downward when jet2 was moving upward along the same flux tube, jet1 may bring downward gas pressure to jet2, 
which act as an extra force that decelerates jet2 \citep{Huang2020}. 
Moreover, the flux tube was filled with hot and dense plasma from jet1, which exerts a viscous drag force during the fast propagation of jet2 \citep{vrs01}.
In contrast, the accelerations in the ascending and descending phases of jet3 are equivalent and lower than $g_{\odot}$ (see Table~\ref{Tab2}), implying that extra forces are absent during jet3.
Therefore, the inconsistency in the accelerations of jet2 is most probably caused by the influence of jet1.
State-of-the-art numerical simulations are needed to carry out a in-depth investigation of the dynamics of coronal jets \citep{wyp18}.

\section{Summary} \label{sect:sum}
In this paper, we carry out multiwavelength observations of three recurring jets on 2014 November 7. 
The jets originated from the same region at the edge of AR 12205 and propagated along the same coronal loop.
The eruptions were generated by magnetic reconnection, which is evidenced by continuous magnetic cancellation at the jet base.
The projected initial velocity of the jet2 is $\sim$402 km s$^{-1}$. The accelerations in the ascending and descending phases of jet2 are not consistent, the former is considerably larger than 
the value of $g_{\odot}$ at the solar surface, while the latter is lower than $g_{\odot}$. There are two possible candidates of extra forces acting on jet2 during its propagation. 
One is the downward gas pressure from jet1 when it falls back and meets with jet2. The other is the aerodynamic drag force during the fast propagation of jet2.
As a contrast, the accelerations of jet3 in the rising and falling phases are constant, implying that the propagation of jet3 is not significantly influenced by extra forces.

\begin{acknowledgements}
The authors are grateful to the colleagues in Purple Mountain Observatory for their constructive suggestions and comments.
SDO is a mission of NASA\rq{}s Living With a Star Program. AIA and HMI data are courtesy of the NASA/SDO science teams. 
This work is supported by the NSFC grants (No. 11790302, 11790300, 11773079), CAS Key Laboratory of Solar Activity, National Astronomical Observatories (KLSA202006), 
and the Strategic Priority Research Program on Space Science, CAS (XDA15052200, XDA15320301).
\end{acknowledgements}

\bibliographystyle{raa}
\bibliography{bibtex}

\label{lastpage}

\end{document}